\documentclass{article}

    \usepackage[final]{neurips_2022}

\usepackage[utf8]{inputenc} 
\usepackage[T1]{fontenc}    
\usepackage{hyperref}       
\usepackage{url}            
\usepackage{booktabs}       
\usepackage{amsfonts}       
\usepackage{nicefrac}       
\usepackage{microtype}      
\usepackage{xcolor}         
\usepackage{chemformula}

\usepackage[sorting=none]{biblatex}
\usepackage{tabularx}
\newcolumntype{A}{>{\centering}p{0.12\textwidth}}
\newcolumntype{B}{>{\centering}p{0.09\textwidth}}
\usepackage{graphicx}
\usepackage{amsmath}
\DeclareMathOperator*{\argmax}{argmax}
\usepackage{subfigure}

\title{Deep Reinforcement Learning for Inverse Inorganic Materials Design}

\author{
Elton Pan\thanks{Equal contribution.}, $ $ Christopher Karpovich,$^\ast$ Elsa Olivetti \\
  Department of Materials Science and Engineering\\
  Massachusetts Institute of Technology\\
  Cambridge, MA 02139, USA \\
  \texttt{\{eltonpan, ckarp, elsao\}@mit.edu} \\
  }
\begin{document}

\maketitle

\begin{abstract}
  A major obstacle to the realization of novel inorganic materials with desirable properties is the inability to perform efficient optimization across both materials properties and synthesis of those materials. In this work, we propose a reinforcement learning (RL) approach to inverse inorganic materials design, which can identify promising compounds with specified properties and synthesizability constraints. Our model learns chemical guidelines such as charge and electronegativity neutrality while maintaining chemical diversity and uniqueness. We demonstrate a multi-objective RL approach, which can generate novel compounds with targeted materials properties including formation energy and bulk/shear modulus alongside a lower sintering temperature synthesis objectives. Using this approach, the model can predict promising compounds of interest, while suggesting an optimized chemical design space for inorganic materials discovery.
\end{abstract}

\section{Introduction}
The discovery of new materials with desirable properties is crucial to applications in areas such as energy storage and electronics manufacturing. However, virtual materials screening is still gated by time- and resource-consuming physics-based simulations and high-throughput experimentation. In recent years, machine learning (ML) has revolutionized inverse materials design, where given one or more target properties, a model is trained to generate candidate compounds that satisfy the desired constraints. Generative neural networks such as autoencoders have been used to optimize organic molecules with desirable properties over a learned latent space \cite{gomez2018automatic, lim2018molecular}. However, optimizing over a non-convex objective function over a high-dimensional latent space is difficult \cite{zhou2019optimization}. Generative adversarial networks (GANs) have also seen success in molecular generation tasks. While they circumvent the need for latent space optimization, GANs suffer from issues such as training instability and mode collapse, which make it difficult to generate compounds with high diversity and validity \cite{sanchez2017optimizing}. RL approaches to molecular generation optimize molecules with respect to a reward function. Prior work includes both policy-gradient and Q-learning approaches to optimization of \textit{organic} molecules and has demonstrated these methods to be sample efficient, stable during training, and high performance \cite{popova2018deep, zhou2019optimization, olivecrona2017molecular}. To the best of our knowledge, no prior work has leveraged RL for \textit{inorganic} materials generation. Here, we propose a deep Q-network (DQN) approach for inorganic material composition generation. Our main contributions include:


\begin{itemize}
  \item First known demonstration of RL for inorganic materials generation
  \item Multi-objective optimization of both inorganic materials synthesis and property constraints
  \item Generation of novel, chemically diverse inorganic compounds with desirable properties
\end{itemize}




\begin{figure}[t!]
    \centering
    \subfigure(a){\includegraphics[scale=0.28]{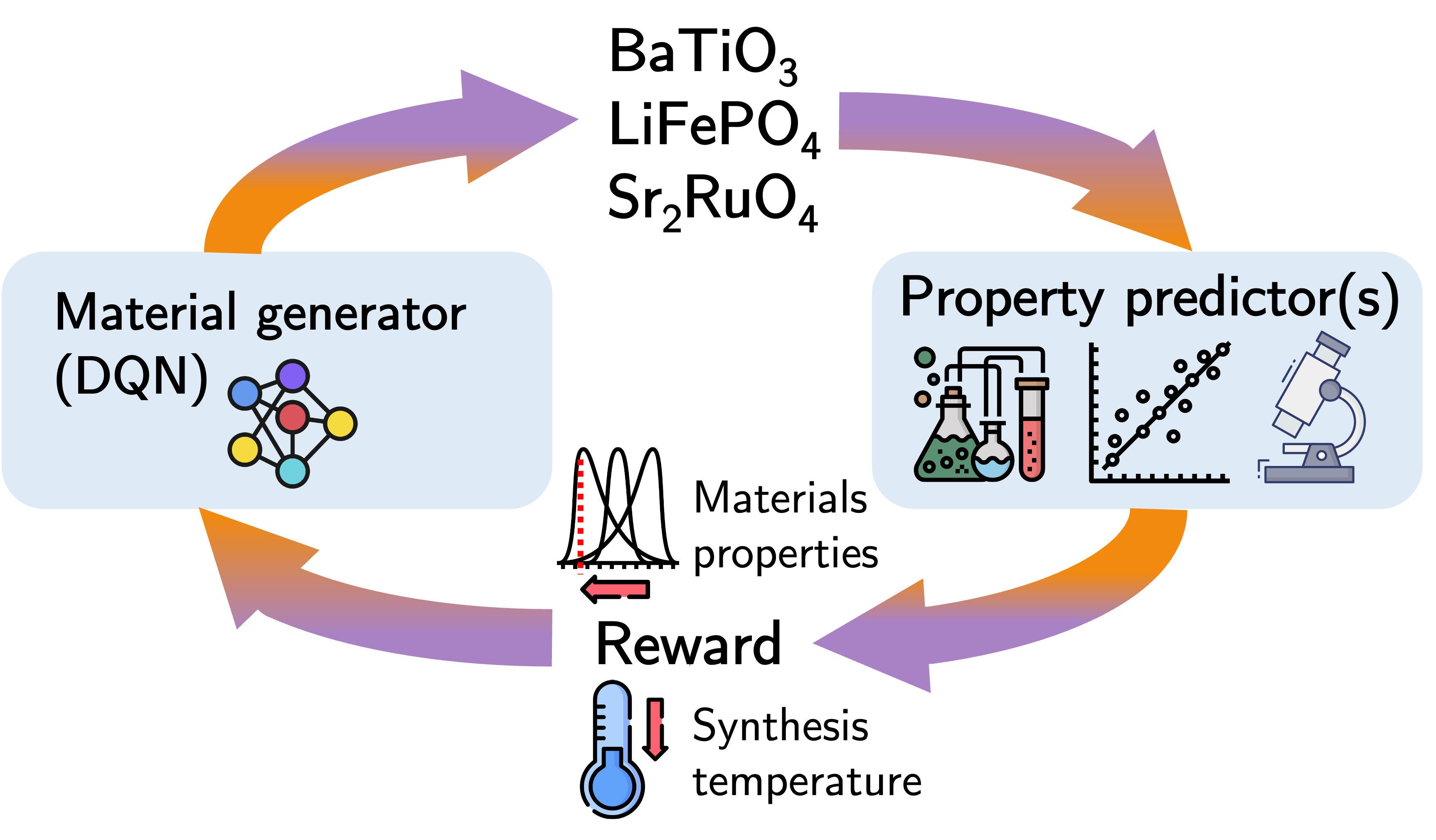}}
    \subfigure(b){\includegraphics[scale=0.31]{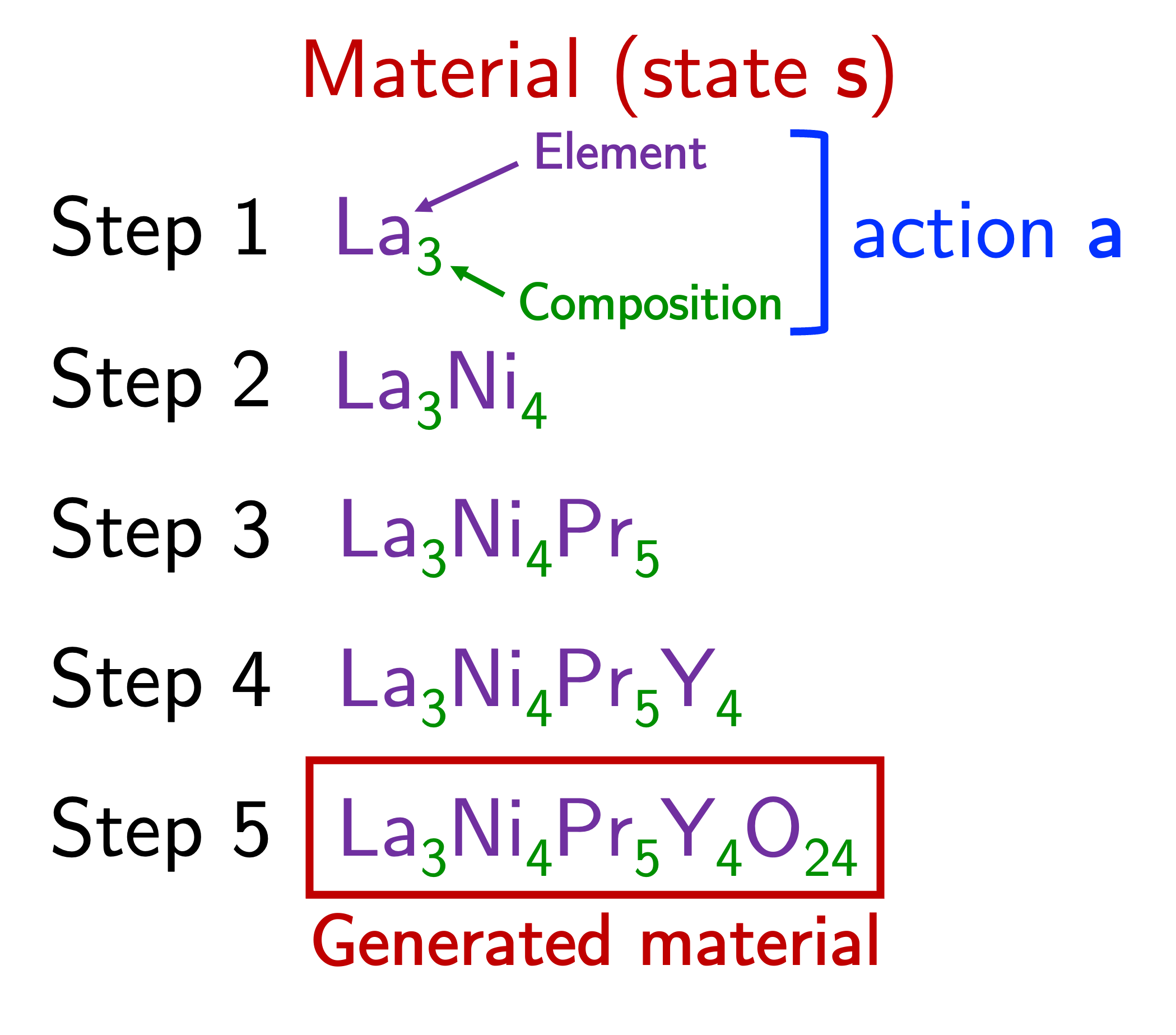}}
    \caption{Schematic of (a) deep Q-network for material design where agent is assigned rewards based on predicted material properties in a feedback loop (b) material generation process. At each step of the generation sequence, state $\mathbf{s}$ is represented by material composition while action $\mathbf{a}$ constitutes the addition of an element (eg. La in step 1) and its corresponding composition (eg. 3 in step 1).}
    \label{RL_loop_material_generation}
\vspace{-3.0mm}
\end{figure}

\section{Methods}
\subsection{Deep Q-Network for materials design}
Q-learning is a model-free reinforcement learning algorithm which trains an agent to behave optimally in a Markov process \cite{watkins1992q}. The agent performs actions to maximize the expected sum of all future discounted rewards given an objective function $J=\sum_{t=0}^{t_f} \gamma^{t} R_{t}\left(\mathbf{s}_{t}, \mathbf{a}_{t}\right)$
where $ \gamma \in[0,1]$ is the discount factor and $R_t$ represents the reward at step $t$ given values $\mathbf{s}_t$ and $\mathbf{a}_t$. For a policy $\pi$ an action-value function can be defined as
\begin{equation}
Q^{\pi}(\mathbf{s}_t, \mathbf{a}_t)={R}_{t+1}+\gamma \max_{\mathbf{a}_{t+1}} Q^{\pi}\left(\mathbf{s}_{t+1}, \mathbf{a}_{t+1}\right)
\label{Bellman}
\end{equation}
with $Q^{\pi}\left(\mathbf{s}_{t+1}, \mathbf{a}_{t+1}\right)$ being the expected sum of all the future rewards the agent receives in the resultant state $\mathbf{s}_{t+1}$. 

In the context of materials design, the agent uses a policy $\pi$ to maximize the expected future reward (material property) via a feedback loop by learning a deep Q-network (DQN) $Q^\pi$ as shown in Fig. \ref{RL_loop_material_generation}(a). The agent (material generator) generates materials, which are then evaluated by material property predictor(s), hence assigning the reward $R$ to the agent. Here, we formulate the material generation process as a \textit{sequence generation} task 
$\mathbf{s_0},\mathbf{a_0},R_0,\mathbf{s_1},\mathbf{a_1},R_1,...$ as shown in Fig. \ref{RL_loop_material_generation}(b) using DQN, where states $\mathbf{s}$ are represented by material compositions while actions $\mathbf{a}$ constitute the addition of an element and its corresponding composition to the existing incomplete material. The goal here is to learn a DQN material generator that maximizes expected rewards to generate compounds of desired properties. More details can be found in Appendices \ref{Q-learning}. and \ref{Material generation}.
\vspace{-2.0mm}

\subsection{Constraint satisfaction for chemical validity}
\label{Constraint satisfaction for chemical validity}
Without contraints a DQN may generate invalid materials through unbounded exploration. To address this, charge neutrality and electronegativity balance constraints are incorporated using a constraint oracle \cite{pan2021constrained} (a trained function) $\hat{c}_{j, t}$ which is formulated as
\begin{equation}
    \hat{c}_{j, t} = \max_{t' \geq t}\left[c_{j, t'}\right] 
    \label{oracle}
\end{equation}
with $c_{j, t}$ being the $j$th constraint to be satisfied at step $t$, and the $\hat{c}_{j, t}$ is determined by the maximum level of violation to occur in all current and future steps $t'$ in the generation process. The constraint oracles $C_{j}(\mathbf{x}, 
\mathbf{a})$ are learned functions (trained on $\hat{c}_{j, t}$) of the probability of an action satisfying a constraint in all future steps. Consequently, the \textit{constrained} agent now follows a \textit{modified} policy
\vspace{-0.5mm}
\begin{equation}
    \pi(\mathbf{s}) = \argmax_\mathbf{a} \biggl[ Q^\pi(\mathbf{s}, \mathbf{a})+ \infty\cdot \min\biggl(0,C_{j}(\mathbf{s}, \mathbf{a})-T \biggr) \biggr] 
\label{modified_policy}
\end{equation}
where $T\in[0,1]$ is a threshold set at 0.5. This important modification to DQN ensures that actions predicted (by constraint oracle $C_{j}(\mathbf{s}, \mathbf{a})$) to generate invalid compounds are penalized by the last term in Eq. \ref{modified_policy}, increasing the probability of generating chemically valid compounds. 

\subsection{Probabilistic sampling of actions for chemical diversity}
\label{Probabilistic sampling of actions for chemical diversity}
DQN agents typically take the greedy action that maximizes the expected return, rendering the policy deterministic in a static environment. However, for a materials generation task, a deterministic policy would be problematic as the agent would \textit{always} take the same actions and generate the \textit{same} material over and over again. This would not be optimal as materials generated by the agent must be chemically \textit{diverse}. To address this, we implement another crucial modification to DQN. Instead of taking the greedy action, the policy $\pi$ is given where the action is \textit{sampled} from the top $n\%$ of actions such that 
$\pi(\mathbf{s}) = \text{random.sample}(\mathbf{a}_1^*,\dots, \mathbf{a}_n^*)$
where $\mathbf{a}_1^*,\dots, \mathbf{a}_n^*$ refer to the top $n\%$ of actions ranked by the $Q^\pi$ and constraint oracles $C_{j}(\mathbf{s}, \mathbf{a})$. Related sampling strategies include Boltzmann sampling \cite{watkins1989models}. Consequently, this modification ensures that the DQN takes actions in a stochastic fashion, which enables the generation of diverse materials. 
\vspace{-2.0mm}

\begin{table}[t]
\caption{Validity, diversity and properties of materials generated by DQN vs. baseline. \textbf{Tasks.} Form: formation energy, Bulk: bulk modulus, Shear: shear modulus, Sinter: sintering temperature, Random: random baseline. \textbf{Metrics.} Char. neut.: charge neutral, Elec. bal.: electronegativity balanced. Values and subscripts are mean and std of $10^3$ generated materials, respectively.}
\begin{tabularx}{\textwidth}{Acccccc}
\toprule
& \multicolumn{2}{c}{\textbf{Validity}}   & \multicolumn{2}{c}{\textbf{Diversity}}  & \multicolumn{2}{c}{\textbf{Material property}}      \\
\cmidrule(lr{1em}){2-3}                     \cmidrule(lr{1em}){4-5}                   \cmidrule(lr{1em}){6-7} 
Task & \% Char. neut. &  \% Elec. bal.  & \% Unique & ElMD & Property 1 & Property 2 \\
     & ($\uparrow$) &  ($\uparrow$) & ($\uparrow$) & ($\uparrow$) & (in Task) & (in Task) \\
\midrule
Form ($\downarrow$)                  &  41.0  & 60.1  & 100  & 14.1$_{7.6}$  & \textbf{-2.54$_{0.72}$}  & -       \\   
Bulk ($\uparrow$)                    &  \textbf{85.3}  & \textbf{65.8}  & 100  & 7.5$_{4.0}$  & \textbf{5.35$_{0.21}$}  & -       \\   
Shear ($\uparrow$)                   &  \textbf{82.8}  & \textbf{69.2}  & 100  & 11.0$_{5.0}$  & \textbf{4.46$_{0.40}$}  & -       \\   
Sinter ($\downarrow$)                 &  \textbf{85.7}  & \textbf{64.7}  & 100  & 12.1$_{7.4}$  & \textbf{827$_{98}$}  & -       \\   
Sinter+Bulk ($\lambda$ = 62.5)   &  85.9  & 61.6  & 100  & 11.1$_{6.2}$  & 826$_{103}$  &    4.24$_{0.77}$    \\
Sinter+Bulk ($\lambda$ = 125)   &  84.7  &  62.8 & 100  & 10.1$_{5.0}$   & \textbf{886$_{102}$}  &    \textbf{4.73$_{0.67}$}        \\
Sinter+Bulk \\ ($\lambda$ = 250)   &  92.6  & 63.9  & 100  & 8.8$_{4.4}$    & \textbf{935$_{64}$}  &    \textbf{5.22$_{0.36}$}          \\

\midrule
Random                     &  59.9  &  42.8 & 100  & 20.1$_{9.8}$  &   \multicolumn{2}{c}{-1.37$_{0.73}$ (Form) }        \\  
(baseline)                &    &   &  &   &   \multicolumn{2}{c}{4.29$_{0.60}$ (Bulk)} \\  
                          &    &   &  &   &   \multicolumn{2}{c}{3.65$_{0.72}$ (Shear)} \\ 
                          &    &   &  &   &   \multicolumn{2}{c}{1011$_{131}$ (Sinter)} \\ 
\toprule
\end{tabularx}
\label{metrics_table}
\vspace{-3mm}
\end{table}

\section{Experiments}
\vspace{-1.0mm}
We evaluate the DQN agents based on metrics related to real-world material discovery process. These metrics relate to stability and validity of compounds: formation energy per atom, charge neutrality, electronegativity balance \cite{dan2020generative}. In addition, performance-based material properties of interest include bulk modulus and shear modulus, which are measures of mechanical strength of a material. Since synthesis also plays a crucial role in real-world materials discovery, sintering temperature (a key synthesis parameter) of material is included. Finally, to evaluate the diversity of compounds generated, the element movers distance (ElMD) \cite{hargreaves2020earth} and $\%$ uniqueness are employed.

The DQN agents are evaluated based on generating materials for 1) \textbf{single-property} 2) \textbf{multi-property} optimization. 
\vspace{-1.5mm}


\begin{figure}[t]
    \makebox[\textwidth][c]{\includegraphics[width=1.25\textwidth]{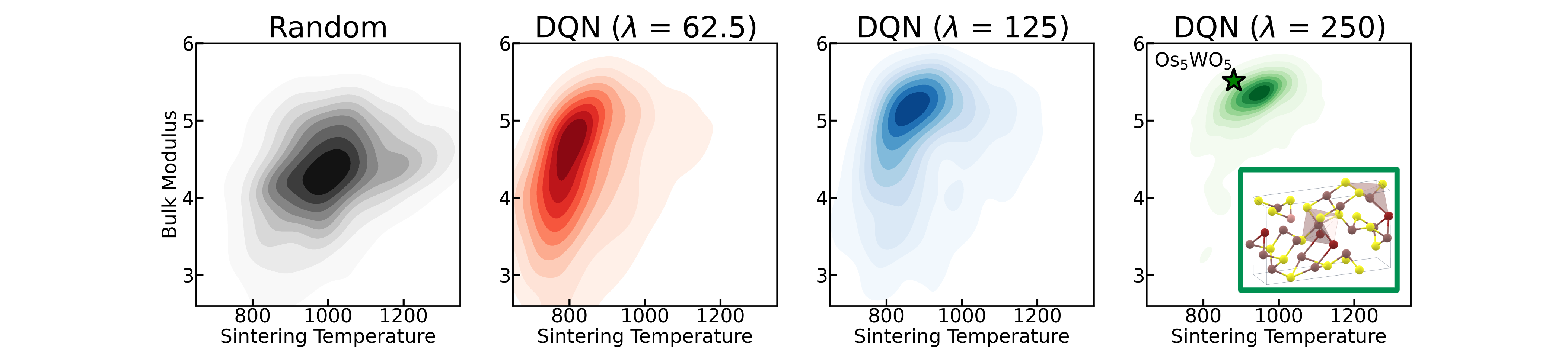}}
  \caption{Property distributions of materials generated by DQN vs. baseline for the multi-property optimization task (\textit{minimize} sintering temperature \textbf{and} \textit{maximize} bulk modulus). DQN generates materials that have \textit{simultaneous} improvements over the baseline. As $\lambda$ 
  increases, DQN prioritizes generating materials with high bulk modulus over those with low sintering temperature. Green star: \ch{Os5WO5}, a \textit{novel} DQN-generated Pareto-optimal material. Green box: its predicted crystal structure.}
  \label{fig:joint_optimization_kde}
\vspace{-3mm}
\end{figure}

\subsection{Single-property material optimization}
\vspace{-0.5mm}
\textbf{Setup.} The task is to generate materials that optimize for a \textit{single} property of interest. We generate $10^3$ materials using the aforementioned method with hyperparameters presented in Appendix \ref{Material generation}. Subsequently, the properties of the generated materials are predicted using trained ML property predictor(s), and their charge neutrality and electronegativity balance are checked via the procedure in \cite{dan2020generative}. Details on dataset curation and ML property predictor training can be found in Appendices \ref{preprocessing} and \ref{predictor}, respectively. We report the percentage of generated materials that satisfy each validity constraint. Here, the goal is to: \textbf{1)} Minimize formation energy per atom: to maximize stability of material \textbf{2)} Maximize bulk modulus and shear modulus: to maximize mechanical strength of material \textbf{3)} Minimize sintering temperature: lower temperatures are associated with lower energy consumption costs and greener synthesis pathways.




\textbf{Results.} For all tasks except for Form in Table \ref{metrics_table}, the DQN outperforms the baseline in terms of generating materials that are charge neutral and electronegativity balanced. This is due to the constraints enforced by the constraint oracles as described in Section \ref{Constraint satisfaction for chemical validity}, resulting in generation of valid materials. For uniqueness, the sampling strategy in Section \ref{Probabilistic sampling of actions for chemical diversity} allows DQN to achieve 100$\%$ uniqueness. 
According to material properties in Table \ref{metrics_table}, DQN generates materials with significantly improved properties over the baseline: more negative formation energy, higher bulk/shear modulus and lower sintering temperature. Generally, the standard deviation (see subscripts) of DQN-generated material properties are lower vs. the baseline, which is  supported by the lower ElMD for DQN. This could be explained by DQN prioritizing exploitation over exploration of chemical space, and is testament to the inherent trade-off between chemical diversity and optimal material property.


\vspace{-1.0mm}

\subsection{Multi-property material optimization}
\vspace{-0.5mm}
\textbf{Setup.} The task is to generate materials that jointly optimize for multiple properties of interest, which is more challenging and more closely represents real-world materials design. The generation protocol is similar to that of single-property task. Here, the goal is to \textit{jointly} maximize bulk modulus and minimize sintering temperature. The reward function is now a linear combination $R = P_1 + \lambda P_2$
where $P_1$ and $P_2$ are rewards corresponding to individual material properties. $\lambda$ is a hyperparameter to weigh the relative importance of individual material properties in the joint optimization. Here, we investigate the influence of varying $\lambda$ on the properties of generated materials. The evaluation metrics are similar to that of the single-property task.

\textbf{Results.} In Fig. \ref{fig:joint_optimization_kde}, as $\lambda$ increases, the DQN prioritizes maximizing bulk modulus over minimizing sintering temperature (shifts up and right). When $\lambda = 250$ or $125$, DQN generates materials that have simultaneous improvements in both material properties over the baseline (Table \ref{metrics_table}) as shown in the shift to the upper-left corner of the subplots in Fig. \ref{fig:joint_optimization_kde}. Moreover, Table \ref{metrics_table} clearly shows that DQN-materials have higher rates of validity (Char. neut. and Elec. bal.) but with lower diversity (ElMD). However, when $\lambda = 62.5$, it over-prioritizes minimizing sintering temperature at the expense of bulk modulus (Table \ref{metrics_table}). This shows the trade-off between the two properties, resulting in a characteristic Pareto front in multi-objective optimization, where $\lambda$ is a parameter that could be tuned according to the application. We also show a \textit{novel}, DQN-generated Pareto-efficient material \ch{Os5WO5} (green star in Fig. \ref{fig:joint_optimization_kde}) with its crystal structure (space group: $Pmn2_1$) predicted by a template matching algorithm \cite{wei2022tcsp}. This approach has enabled us to suggest materials, but to assess them and understand their stability future work must involve validation using simulation-based quantum chemical methods.

\section{Broader Impact}
The ability to propose \textit{novel}, \textit{chemically valid} compounds with \textit{desirable properties} in terms of synthesis and property constraints is an important step towards bridging experiment and computation to accelerate AI-guided materials design for tackling pressing modern-day
challenges.

\printbibliography

\appendix

\section{Appendix}

\subsection{Data acquisition and preprocessing}
\label{preprocessing}
\textbf{Datasets.}
To train the materials property predictor models, we leveraged a subset of inorganic materials and their computed properties contained within the Materials Project (MP) database \cite{jain2013commentary}. Materials Project is a widely used inorganic materials database containing crystal structures and materials properties data calculated from high-throughput quantum mechanical calculations. To train the sintering temperature predictor model, we used a publicly released inorganic solid-state synthesis database text-mined from scientific literature using a combination of NLP and rule-based extraction techniques \cite{kononova2019text}.

To preprocess the materials property data, we use a similar preprocessing strategy as \cite{jha2018elemnet, dan2020generative, pathak2020deep}. For a formula with multiple reported formation energies, we choose the lowest one to select the most stable compound. We additionally removed all single-element compounds and any compounds with a formation energy outside the interval $[\mu-5\sigma, \mu+5\sigma]$, where $\mu$ and $\sigma$ are the average and standard deviation of the formation energies of all compounds in the MP dataset. For each property, all entries which did not contain a computed property value were discarded. After preprocessing, we obtained datasets containing the formation energies for 68,708 compounds as well as the bulk and shear modulus values for 9,888 compounds. For bulk and shear moduli, property values were taken as their corresponding logarithm.

To preprocess the extracted sintering temperature data, reactions without at least one sintering step with a reported temperature were removed. Temperatures were converted into units of Celsius ($^{\circ}$C) and limited to between 200 $^{\circ}$C and 2000 $^{\circ}$C. If sintering occurred more than once in a recipe, the last operation in chronological order was taken. For reactions with more than one reported occurrence in literature (same target and precursors), the ground truth reaction condition was taken to be the average of the reported conditions. From this dataset, we extracted the experimental sintering temperatures for 12,296 inorganic compounds.

\subsection{ML property prediction models}
\label{predictor}
We built ML property prediction models for both materials properties (formation energy, bulk/shear modulus) and synthesis objectives (sintering temperature). For the materials property prediction, we leveraged Roost \cite{goodall2020predicting}, a graph-composition message passing neural network architecture which can predict materials properties from composition. Models were trained based on default hyperparameters presented in package using a random 90/10 train-test split.

For sintering temperature prediction, inorganic compounds were featurized using Magpie compositional features, which are physically motivated descriptors that take the form of a 145-dimensional embedding containing stoichiometric properties, elemental properties, electronic structure, and ionic compound features \cite{ward2016general}. To predict sintering temperature from materials composition, we trained a random forest (RF) model and optimized hyperparameters using a 5-fold cross-validation. 

Metrics of these prediction models are reported in Table \ref{ML_property_predictor_metrics_table}.

\begin{table}[ht]
\centering
\caption{Performance metrics of ML property prediction models.}
\begin{tabularx}{13.25cm}{cccccc}
\toprule
Task & Model &  Dataset  & Test $R^2$ & Test MAE & Test RMSE \\
\midrule
Formation energy per atom (eV) & Roost &  MP  & 0.913 & 0.1367 & 0.3467 \\
Bulk modulus (log MPa)             & Roost &  MP  & 0.753 & 0.2331 & 0.4323 \\
Shear modulus (log MPa)            & Roost &  MP  & 0.701 & 0.3321 & 0.5135 \\
Sintering temperature ($^{\circ}$C)    & RF    &  MP  & 0.854 & 52.09  & 96.63 \\
\toprule
\end{tabularx}
\label{ML_property_predictor_metrics_table}
\end{table}

\subsection{Deep Q-Network}
\label{Q-learning}
In Q-learning (or DQN), the $Q$-value is the expected discounted reward for a given state and action, and therefore the optimal policy $\pi^*$ can be found using iterative updates with the Bellman equation (Eq. (\ref{Bellman})). Upon convergence, the optimal $Q$-value $Q^*$ is defined as:
\begin{equation}
Q^{*}\left(\mathbf{s}_t, \mathbf{a}_t\right)=\mathbb{E}_{\mathbf{s}_{t+1} \sim p}\left[R_{t+1}+\gamma \max _{\mathbf{a}_{t+1}} Q^{*}\left(\mathbf{s}_{t+1}, \mathbf{a}_{t+1}\right) \bigm| \mathbf{s}_t, \mathbf{a}_t\right]
\end{equation}
$Q\left(\mathbf{s}_t, \mathbf{a}_t\right)$ can be represented by a function approximator.
The Q-function is approximated with a deep Q-network (DQN) $Q_\theta$ parameterized by weights $\theta$ \cite{mnih2013playing}. 

\begin{figure}[ht]
    \centering
    \includegraphics[scale=0.35]{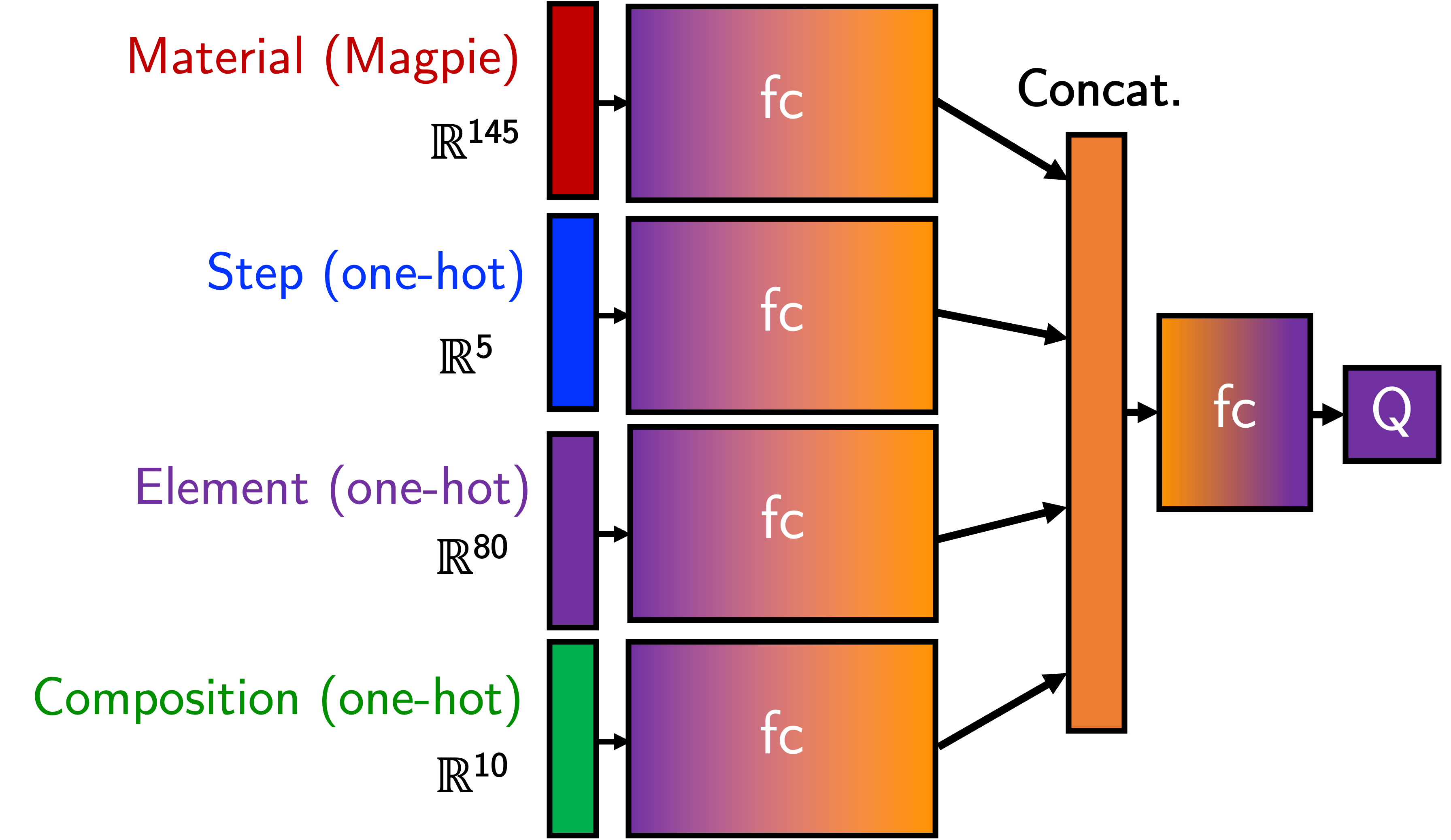}
    \caption{Architecture of deep Q-network.}
    \label{fig:dqn_architecture}
\end{figure}

\textbf{DQN architecture.} The inputs are state $\mathbf{x}_t$ and action $\mathbf{a}_t$ at the corresponding step $t$. The architecture of the DQN can be found in Fig. \ref{fig:dqn_architecture}. Leaky ReLU is used in all fully connected (fc) layers. The 4 vectors (vide infra) are then individually passed through fc layers, and their corresponding hidden states ($dim = 64$) are concatenated into a single vector ($dim = 256$) before passing through another fc layer to give the Q-value.

The state representation has two components:
\begin{enumerate}
    \item Material composition: Each generated material (complete or incomplete) is featurized using Magpie to give a $\mathbb{R}^{145}$ vector as described in A.\ref{ML_property_predictor_metrics_table}.
    \item Step: Step of generation. It has been reported in previous works \cite{zhou2019optimization} that the inclusion of the step (time-dependent policy) outperforms exclusion of such (time-independent policy). Since horizon is capped at 5 steps, this is represented by a one-hot encoding of $\mathbb{R}^{5}$.
\end{enumerate} 

The action representation has two components:
\begin{enumerate}
    \item Element: The atomic species being added to the sequence. Since a element set consists of 80 elements, this is represented by a one-hot encoding of $\mathbb{R}^{80}$.
    \item Composition: The numerical subscript of the atomic species that is added to the sequence. Since this ranges from 0-9, this is represented by a one-hot encoding of $\mathbb{R}^{10}$.
\end{enumerate}

\textbf{DQN training.} The DQN is trained over 500 iterations with the use of a replay buffer of size 50000 that addresses the issue of correlated sequential samples \cite{lin1993reinforcement}. In each iteration, 100 compounds were generated, stored in the replay buffer, and Q-network is trained on 100 samples randomly sampled from the replay buffer with smooth L1 loss function and Adam optimizer with learning rate of 0.01. Initial exploration is encouraged using an $\epsilon$-greedy policy starting with high $\epsilon$ value of 0.99, which is decayed (by a factor of 0.99 after each iteration) over the course of training to ensure eventual exploitation and convergence to the optimal policy. Discount factor $\gamma$ is set at 0.9. Models are trained on a single NVIDIA RTX A5000 GPU. Training a DQN agent generally takes 2-3 hours.


\begin{table}[h]
\centering
\caption{Reward functions of optimization tasks. $E$ = formation energy per atom, $K$ = bulk modulus, $G$ = shear modulus, $S$ = sintering temperature, $\lambda$ = weighing coefficient to weigh individual rewards.}
\begin{tabularx}{8cm}{cc}
\toprule
Task & Reward \\
\midrule
Formation energy per atom & $-E$  \\
Bulk modulus              & $K$  \\
Shear modulus             & $G$  \\
Sintering temperature     & $-S$     \\
Sintering temperature + Bulk modulus       & $-S + \lambda K$     \\
\toprule
\end{tabularx}
\label{rewards_table}
\end{table}

\subsection{Material generation}
\label{Material generation}
The material generation process is formulated as a \textit{sequence generation} task 
$\mathbf{s_0},\mathbf{a_0},R_0,\mathbf{s_1},\mathbf{a_1},R_1,...$ as shown in Fig. \ref{RL_loop_material_generation}(b) with a horizon capped at 5 steps. At each step, an element (out of a set of 80 possible elements) and its corresponding composition (0-9) are appended to the existing (or empty) material. If the corresponding composition is 0, the agent does not add the element. This allows the agent to generate materials of any length of $\leq$ 5 elements (length of horizon) hence increasing chemical diversity. In all experiments in this work, we restrict to generating oxides, hence any element can be added in the first 4 steps, but only oxygen for the final step.

To increase the diversity of generated materials, the action is \textit{sampled} from the top $n\%$ of actions ranked by the $Q^\pi$ and constraint oracles $C_{j}(\mathbf{s}, \mathbf{a})$. Empirically, it was found that $n$ can be increased from 0 up until 20 without a deterioration of material properties, hence $n$ was set to 20 to maximize diversity of generated materials. Any larger value than $n$ = 20 would further improve diversity, but at the expense of material properties. Rewards at all non-terminal are assigned zero until step 5 (terminal step) where the reward $R$ is assigned according to Table \ref{rewards_table}.

\end{document}